\documentclass[nofootinbib,superscriptaddress,10pt]{revtex4}

\usepackage[pdftex,bookmarks=true,linktocpage=true]{hyperref}
\usepackage[a4paper,lmargin=70pt,rmargin=70pt]{geometry}
\usepackage{amsmath}
\usepackage{amssymb}
\usepackage{slashbox}
\usepackage{graphicx}
\newcommand{\ed}{\,.}
\newcommand{\ec}{\,,}
\newcommand{\ecq}{\ec\quad}

\newcommand{\cO}{\ensuremath{\mathcal{O}}}
\newcommand{\Pf}{\ensuremath{\text{Pf}}}
\newcommand{\parts}{\ensuremath{\text{particles}}}
\newcommand{\SO}{\ensuremath{\text{SO}}}
\newcommand{\pJ}{\ensuremath{p_J}}
\newcommand{\mJ}{\ensuremath{m_J}}
\newcommand{\en}{\ensuremath{\varepsilon}}
\DeclareMathOperator{\trace}{Tr}

\begin{document}
\begin{flushright}
CERN-PH-TH/2011-007
\end{flushright}

\title{Classification of Energy Flow Observables in Narrow Jets}
\author{Guy Gur-Ari} \address{Department of Particle Physics \&
Astrophysics, Weizmann Institute of Science, Rehovot 76100, Israel}

\author{Michele Papucci}
\address{CERN, PH-TH, CH-1211, Geneva 23, Switzerland}
\address{on leave from Lawrence Berkeley National Laboratory, Berkeley, CA 94720}

\author{Gilad Perez} \address{Department of Particle Physics \&
Astrophysics, Weizmann Institute of Science, Rehovot 76100, Israel}

\setlength{\parindent}{0pt}
\setlength{\parskip}{1ex}

\begin{abstract} 
  We present a classification of energy flow variables for highly
  collimated jets. Observables are constructed by taking moments of the
  energy flow and forming scalars of a suitable Lorentz subgroup. The jet
  shapes are naturally arranged in an expansion in both angular and energy
  resolution, allowing us to derive the natural observables for
  describing an $N$-particle jet. We classify the leading variables that
  characterize jets with up to 4 particles. We rediscover the familiar
  jet mass, angularities, and planar flow, which dominate the lowest
  order substructure variables. We also discover several new observables
  and we briefly discuss their physical interpretation.
\end{abstract}

\maketitle

\section{Introduction}

Analyzing the substructure of highly-boosted massive jets ($\pJ \gtrsim 500$ GeV)
has proved to be useful for distinguishing new physics signals from the
QCD background. For instance, jets originating from boosted electroweak gauge bosons~\cite{Butterworth:2002tt}, tops~\cite{KKG1,KKG2}, Higgs~\cite{Butterworth:2008iy}, 
and even new physics particles~\cite{Butterworth:2009qa,Butterworthsparticles} 
are all interesting targets of searches
conducted at the Tevatron and the LHC experiments, and it is therefore important to be able to
distinguish them from QCD jets. For recent reviews on substructure
techniques, experimental status and new physics searches see~\cite{Abdesselam:2010pt} and references therein.

One way to characterize jet substructure is to consider observables which
are functions of the energy flow within the jet, namely the energy distribution
$\en(\vec{x})$ as measured by the detector, where $\vec{x}$ are the coordinates
on the detector surface. Examples of such observables include the jet
mass and the angularities \cite{Berger:2004xf}\footnote{Here we use a
different normalization than that of \cite{Berger:2004xf}.}, 
\begin{align}
  \tau_a &\equiv \frac{1}{E_J} \sum_{i \in \parts} 
  E_i \, (\sin \theta_i)^a \, (1 - \left|\cos\theta_i\right|)^{1-a} \ec
  \label{angularity}
\end{align}
where $E_J$ is the jet energy, $E_i$ is the
energy of particle $i$, $\theta_i$ is the angle relative to the jet axis,
and $a$ is a real parameter restricted to $a \le 2$ to ensure IR safety.
Another example is planar flow \cite{Almeida:2008tp} (see also~\cite{Thaler:2008ju}), which we discuss
below in detail. 

While many observables have been defined, so far there has been no
systematic way to classify them or to construct all observables at a
given order in the detector resolution. The goal of the present work is to
suggest such a classification for the case of narrow jets, as we now
describe.

An energy flow observable needs to meet several criteria.  Ideally, it
should be Lorentz invariant and IR-collinear (IRC) safe. In practice it
is difficult to find observables which are Lorentz
invariant~\cite{Thaler:2008ju}, the jet mass being an exception.  

We may however relax this requirement as follows. For highly-boosted,
narrow jets with fixed 4-momentum the cross section factorizes at leading
order into the hard process and the jet function (in which we are
interested).
Since the jet momentum is an input parameter of the jet function, it is
enough to consider observables which are invariant under the subgroup of
Lorentz that doesn't change the jet momentum. For a massive jet, this
little group is isomorphic to $\SO(3)$, as can be seen by boosting to the
jet rest frame. 

Another property of a jet is that it is contained in a cone of radius $R
\ll 1$ around its axis, and we should further restrict ourselves to
Lorentz transformations that leave the cone
invariant\footnote{Alternatively, we may describe the jet as its
4-momentum plus a detector surface of small area, perpendicular to the
jet axis. The little group should leave the momentum and surface
invariant. Moreover, even if we will refer to the jet cone throughout the paper, our conclusions
are not restricted to jet cone algorithms. For us it is sufficient that there exist a number $\bar R<1$ such that each jet
is contained in cones of radius $\bar R$, which is also generically the case for sufficiently hard jets found with sequential recombination algorithms. 
Indeed, for narrow massive jets, it was shown that the angularity
distributions are similar for a wide range of jet
algorithms~\cite{Almeida:2008yp}.}. The
subgroup that stabilizes both the jet momentum and the cone is $\SO(2)$,
the rotations around the jet axis.


In the present paper we focus on highly-collimated jets, working at
leading order 
in the small cone size $\theta_i \!<\! R \!\ll\! 1$. We will further assume that all the
particles that make up the jet are massless.
Under these assumptions, we propose a complete classification of energy
flow observables which are IR safe and Lorentz invariant, up to Lorentz
transformations that change the jet momentum and the cone. Furthermore
the observables may be naturally arranged as an expansion in terms of the
energy resolution and the cone size (or alternatively the energy and
angular resolutions).

In the narrow cone approximation the experimental calorimeter information of the jet is fully specified by the
energy distribution on a surface perpendicular to the jet axis, $\en(\vec{x})$, with $\vec x\equiv \left(x_1,x_2\right)$ corresponding to a set of coordinates on the two dimensional jet surface. 
To obtain the classification, we begin by describing a given energy
distribution $\en(\vec{x})$ in terms of its moments, 
\begin{align*}
  I_{i_1\dots i_n} &=
  \int d^2x \,\en(x)\, x_{i_1} \cdots x_{i_n}\ed
\end{align*}
Observables are then constructed by taking products of these moments, and
forming $\SO(2)$ scalars by contracting their indices. Invariance under
boosts along the jet axis can be achieved by adding a factor of $E_J$
raised to some power. Since our moments form a straightforward
generalization of angularities with $a \le 2$, they are manifestly IR
safe.

Moments of higher rank are of higher order in the cone size $R$, since $x
\propto \theta$, the opening angle. Moreover, since $I \sim \epsilon(x)$, observables that
are products of a larger number of moments are more sensitive to errors in energy
measurement (we will be more precise about this point below). This allows
us to arrange the observables in an expansion by energy resolution and cone
size as will be shown in section \ref{moment-expansion}.  At the lowest orders in this expansion we find the jet mass and
the angularity with $a=-2$. At the next order we find the angularity with
$a=-4$ and the planar flow. Yet higher orders produce the whole tower of
angularities with $a=-2k$, as well as many new observables.

The paper is organized as follows. In the next section we review the definition of planar flow, showing how
it naturally leads to the expansion of the energy distribution in terms
of moments. In section~\ref{moment-expansion} we define the moment
expansion, find the first few observables, and precisely identify the small
parameters in which we expand. In section~\ref{first-few-orders} we turn
to a classification of the leading observables that characterize jets
with up to 4 particles. In section~\ref{zernike-expansion} we present an alternative approach to the study of 
jet substructure, based on an expansion in terms of orthogonal functions.
Finally, in section~\ref{epsI2I4} we analyze one
of the new observables discovered in the classification, developing
techniques that may be applied to other observables as well. Section~\ref{discussion} contains a few comments
on the application of this formalism to the case of hadron colliders and our conclusions.

\section{Mass, Planar Flow, and the Second Moment}
\label{preliminaries}

Planar flow is defined in terms of the matrix $I_w$, with
components\footnote{We use a different normalization for $I_w$ than that
of \cite{Almeida:2008yp}, without affecting the planar flow definition.}
\begin{align*}
  I_w^{kl} \; &= 
  \sum_{i \in \parts} \!\! E_i
  \frac{p_{i,k}^\perp}{E_i}
  \frac{p_{i,l}^\perp}{E_i} 
  \; \approx
  \sum_{i \in \parts} \!\! E_i
  \, \theta_i f_k(\phi_i)
  \, \theta_i f_l(\phi_i)
  \ec
\end{align*}
where $p_{\perp}$ is the particle momentum in the detector plane, $\phi$
is the azimuthal angle, and 
$
  f_1(\phi) = \cos(\phi) \ec
  f_2(\phi) = \sin(\phi) \ed
$
The approximation is made under the narrow cone assumption.
Planar flow \cite{Almeida:2008yp} is then defined as
\begin{align*}
  \Pf &= \frac{4 \det I_w}{(\trace I_w)^2} \ed
\end{align*}
One can easily verify that $0 \leq \Pf \leq 1$, that it
vanishes when the energy distribution lies on a line, and that it is maximal for
isotropic distribution.

$I_w$ also enters in the definition of the jet mass. From $\mJ^2 = \pJ^2$, $\pJ= \sum_i
p_i$ it is easy to see that, in the narrow cone
approximation, 
\begin{align}
  \frac{\mJ^2}{E_J} \; &\approx
  \sum_{i \in \parts} \!\!\! E_i \theta_i^2 \, = \,
  \trace I_w \ed
  \label{mj}
\end{align}
The narrow cone approximation ensures also that $m_J \ll E_J$.

Now, let $x_k(\theta,\phi) = \theta f_k(\phi)$ denote a coordinate system on the
detector plane. Then we may write $I_w$ as a sum over detector cells,
where we weigh each cell by its total collected energy $E(x)$, namely
\begin{align*}
  I_w^{kl} &= \sum_{n \in \text{cells}} 
  E\left( x^{(n)} \right) x^{(n)}_k x^{(n)}_l \ec\\
  E(x) &= \sum_{i \in \parts} E_i \,
  \delta_{\theta_i,\theta(x)}
  \delta_{\phi_i,\phi(x)}
  \ed
\end{align*}
One can define a normalized version of this tensor,
$I_{kl} = I_w^{kl} / E_J \ec$
where $E_J$ is the jet energy. For the sake of clarity let us rewrite the above expression in
integral form,
\begin{align}
  I_{kl} &= \int d^2x \, \en(x) \, x_k x_l \ec
  \label{I2}
\end{align}
where $\en(x)$ is the continuous, normalized energy distribution, given by
\begin{align*}
  \en(x) &= \frac{1}{E_J} \sum_{n \in \text{cells}} 
  E\left(x^{(n)}\right) \delta(x - x^{(n)}) \ed
\end{align*}
Note that
\begin{align}
  \int d^2 x \, \en(x) = 1 \ed
  \label{moment0}
\end{align}
Equations \eqref{I2} and \eqref{moment0} are quite suggestive.  The
function $\en(x)$ encodes all the information about the jet structure and
it is the object we want to characterize.  Now, we have just seen that
$I_{kl}$ is its second moment and it gives rise to two interesting
physical observables--- the jet mass and the planar flow. It is well
known that all the information of $\en(x)$ is encoded in its moments. It
is therefore plausible that expressing the energy distribution $\en(x)$
in terms of its moments would provide a natural way to derive observables
with the desired properties.  One can also expand the function $\en(x)$
in a set of orthogonal functions and characterize it in terms of its
expansion coefficients. $\SO(2)$ invariance will be ensured by taking
suitable coefficient combinations that are singlets under rotation.  The
two approaches are complementary and can provide  different insight on
the properties of $\en(x)$.  These are the ideas that we will explore in
the remainder of this work.

\section{Expansion in Moments}
\label{moment-expansion}

The $n$-th moment $I_n$ of the energy distribution $\en(x)$ is defined by
\begin{align*}
  I_{k_1,\dots,k_n} &\equiv
  \int d^2x \, \en(x) \, x_{k_1} \cdots x_{k_n} 
  = \frac{1}{E_J}
  \sum_{i\in\parts} \! E_i \, x_{k_1}^{(i)} \cdots x_{k_n}^{(i)}
  \ed
\end{align*}
The zeroth moment is \eqref{moment0}. The first moment is the
expectation value, or dipole. It is set to zero by the requirement that
the total transverse momentum of the jet vanishes, a state which can
be reached by rotating the jet. This fixes the origin of the
detection plane, which in turn determines the jet axis.
We have then
\begin{align*}
  I_0 = 1 \ecq
  I_1 = 0 \ed
\end{align*}

The first non-trivial moment is therefore $I_2$, as expected. We are
looking to define observables that are invariant under the little group
$\SO(2)$, the Lorentz subgroup that doesn't change the jet momentum or
the cone.
$\SO(2)$ has two independent invariant tensors, $\delta_{ij}$ and
$\epsilon_{ij}$. The only $\SO(2)$ scalar that is linear in $I_2$ is
the normalized version of \eqref{mj}, namely
\begin{align*}
  I_{ii} \approx \frac{\mJ^2}{E_J^2} \ed
\end{align*}

Next, consider a tensor product $I_2 \otimes I_2$. There are three
nontrivial scalars one may construct,
\begin{align*}
  I_{ii} I_{jj} \ecq
  I_{ij} I_{ij} \ecq
  \epsilon_{ij} \epsilon_{kl} I_{ik} I_{jl} \ed
\end{align*}
Of these, only two are independent, since
\begin{align*}
  \epsilon_{ij} \epsilon_{kl} I_{ik} I_{jl}
  = 2 ( I_{ii} I_{jj} - I_{ij}^2 ) 
  = 2 \det I \ed
\end{align*}
Also, the first scalar, $I_{ii}I_{jj}$, factorizes in lower-rank scalars.
We therefore find only one new observable, $\det(I)$, which is an un-normalized
version of planar flow.

Before proceeding with the expansion, let us clarify in what sense the
moment expansion is an expansion in small parameters. Planar flow,
composed of $I_2 \otimes I_2$, is apparently of a higher order than the
mass squared, composed of a single $I_2$. To quantify this statement,
first note that energy distribution is constrained to lie
within a small cone of radius $R \ll 1$. Since $I_n \sim x^n \sim
\theta^n$, we have that $I_n$ scales as  $R^n$. As an expansion in $R$, planar
flow is of order 4 while the jet mass squared is of order 2. 

Another small parameter we may consider is the angular resolution
$\Delta\theta$, which is ultimately limited by the calorimeter cell size.  Including its effect gives $I_n \sim (\theta \pm
\Delta\theta)^n = \theta^n \pm n \Delta\theta \,\theta^{n-1} + \cdots$.
The error on the value of an observable due to the finite angular resolution is therefore
proportional to the total rank of moments composing this observable.
Planar flow has error $\sim 4\Delta\theta$, while the mass squared has error
$\sim 2\Delta\theta$. Regardless of which small angular parameter we
choose, measuring a moment of higher rank requires a more accurate
detector.

As for the energy, the only small parameter is the energy resolution
$\Delta\en$. A single moment is proportional to $\en(x)$, so for a given
observable the error due to energy resolution will increase with the
number of moments that make up the observable. Planar flow will tend to have a larger error
than the mass.

\section{The First Few Orders}
\label{first-few-orders}

We are now in a position to classify all the jet energy flow observables.

\begin{table}
  \begin{center}
    \begin{tabular}{r|llll}
      \backslashbox{$R$}{$\Delta\en$} & 1 & 2 & 3 & 4 \\
      \hline
      2 & $I_2$ & - & - & - \\
      4 & $I_4$ & $(I_2)^2$ & - & - \\
      6 & $I_6$ & $I_2I_4,(I_3)^2$ & $(I_2)^3$ & - \\
      8 & $I_8$ & $I_2I_6,I_3I_5,(I_4)^2$ &
          $I_2(I_3)^2,(I_2)^2I_4$ & $(I_2)^4$ 
    \end{tabular}
    \caption{Moment products that correspond to given orders in
    energy resolution $\Delta\en$ and cone size $R$. Observables are
    constructed by contracting these products in various ways.}
    \label{moment-table}
  \end{center}
\end{table}

Table \ref{moment-table} lists the outer products that appear in the
first few orders of the expansion.  Note that there are no observables
with an odd power of $R$ since we cannot fully contract an odd number of
$\SO(2)$ indices. Each outer product may be contracted in several
different ways, giving rise to different observables. However, algebraic relations will reduce the total number of independent contractions.

On top of this expansion it is useful to consider jets that consist of a
given number of particles (or detector cell towers). A jet of $N$ particles is characterized by
$3N - 4$ variables, corresponding to the number of particle momentum
components, minus the jet axis, the jet energy, and the $\SO(2)$ angle
associated to the overall rotation around the jet axis.\footnote{This is at leading order. Soft phenomena, such as color connection with other jets in the event, will generally break this symmetry.} For given $N$ we can
then identify the lowest order, Lorentz-invariant observables that
characterize such jets. Let us list the observables for jets with up to 4
particles.

The two lowest order observables are 
\begin{align*}
  I_{ii} &= \frac{1}{E_J} \sum_{i\in\parts} \!\! E_i \theta_i^2 
  \,\approx \frac{\mJ^2}{E_J^2} \ec\\
  I_{iijj} &= \frac{1}{E_J} \sum_{i\in\parts} \!\! E_i \theta_i^4 
  \,\approx 8\, \tau_{-2} \ed
\end{align*}
Therefore, the natural observables for describing a two-particle jet are
the mass and the angularity $\tau_{-2}$.

To describe 3-particle jets we need 3 more observables in addition to
the mass and $\tau_{-2}$.  As shown above, the outer product $(I_2)^2$ has
one independent contraction, $\epsilon_{ik} \epsilon_{jl} I_{ij} I_{kl}
\sim \det I$, the planar flow. Next we have $I_6$ which has one
contraction, 
\begin{align*}
  I_{iijjkk} &= \frac{1}{E_J} \sum_{i\in\parts} \!\! E_i \theta_i^6 
  \,= 32 \, \tau_{-4}
  \ec
\end{align*}
corresponding to another member of the angularity family with higher $a$.
Generally, $I_{2n}$ with $n>1$ has only one contraction, $I_{i_1i_1\cdots
i_{n}i_{n}} \sim \sum_i E_i \theta_i^{2n}$, which corresponds to an
angularity with $a=2(1-n)$.

At the next order we find $I_2I_4$, $(I_3)^2$, and $I_8$, with the
following independent contractions:
\begin{align*}
  I_2I_4 &: \;
  \epsilon_{ij} \epsilon_{kl} I_{ik} I_{jlmm} \ec\;
  \epsilon_{ij} I_{ik} I_{jkll} \\
  (I_3)^2 &: \;
  \epsilon_{ij} \epsilon_{kl} I_{ikm} I_{jlm} \ec\;
  I_{ijk} I_{ijk} \\
  I_8 &: \; I_{iijjkkll}
\end{align*}
The additional contractions $I_{ij}I_{ijkk}$, $I_{ijj}I_{ikk}$ can be shown to be
linearly dependent on these. Any combination of them can be chosen as
the remaining observable for 3-particle jets. In lumping $I_8$ with the
rest we assumed for simplicity that energy and angular measurements have comparable
weight in the small parameter expansion. Of course, if this assumption is not true, $I_8$
may be preferred or disfavored in comparison with the other contractions.

Note that an observable---such as planar flow---which includes one or
more $\epsilon$ symbols vanishes when all particles are on a
line.\footnote{To see this, first rotate the line configuration to lie on
the $x_1$ axis, and then note that the $\epsilon$ tensor forces an $x_2$
factor to appear. This factor vanishes wherever $\en(x) \ne 0$.} Since
two particles always lie on a line, these observables contribute at
leading order to 3-particle jets. They can therefore be used to
distinguish QCD jets, which favor 2-parton configurations, from
e.g. top jets that favor 3-body decays.

Finally, 4-particle jets are described by eight variables. So far we found
nine leading order observables,
\begin{gather*}
  I_{ii} \ecq
  I_{iijj} \ecq
  \epsilon_{ij} \epsilon_{kl} I_{ik} I_{jl} \ecq
  I_{iijjkk} \ecq
  \epsilon_{ij} \epsilon_{kl} I_{ik} I_{jlmm} \ec \\
  \epsilon_{ij} I_{ik} I_{jkll} \ecq
  \epsilon_{ij} \epsilon_{kl} I_{ikm} I_{jlm} \ecq
  I_{ijk} I_{ijk} \ecq
  I_{iijjkkll}
  \ed
\end{gather*}
Of these, any leading eight can be chosen to describe
the jet.


\section{Expansion in Zernike polynomials}
\label{zernike-expansion}

One can expand $\en(x)$ in a series of orthogonal functions.  Since
$\en(r,\phi)$ is defined on a disc of radius $R$, perhaps the most
convenient expansion is in terms of the Zernike
polynomials~\cite{Zernike}, which form an orthogonal basis on the unit
disc. They are defined by
\begin{gather*}
  Z^m_n(\rho,\phi) = R^m_n(\rho) \cos(m\phi) \ecq
  Z^{-m}_n(\rho,\phi) = R^m_n(\rho) \sin(m\phi) \ecq
  0 \le \rho \le 1 \ec
\end{gather*}
where $0 \le m \le n$, $n-m$ even, and $R_n^m (\rho)$ are a set of
polynomials of degree $n$ respecting the orthogonality condition
\begin{align*}
\int_0^1 {\rm d}\rho \rho R_n^m(\rho) R_{n'}^m(\rho) = \frac{1}{2n+2}\delta_{n,n'} \ed
\end{align*}
The orthogonality among different $m$'s follows trivially from the
orthogonality of the Fourier modes.  This set of functions is widely used
in optics, in particular in the study of optical aberrations where the
expansion coefficients have simple geometrical meaning. 

The expansion of the energy distribution is
\begin{align}
 \en(r,\phi) &= \frac{a_{0,0}}{R^2}  + \frac{1}{R^2}\sum_{n=1}^{\infty} \sum_{\substack{0\leq m\leq n, \\ n-m\,{\rm even}}} \left[ a_{n,m} R_n^m\!\left( \frac{r}{R} \right) \cos(m \phi)+a_{n,-m} R_n^m\!\left( \frac{r}{R} \right)\sin(m \phi) \right] \ec 
 \label{zexpansion}
\end{align}
The conditions $I_0=1$ and $I_1=0$ fix $a_{0,0}=1/\pi$, and $a_{1,\pm1}$
to vanish. One can further expand the moments defined in
section~\ref{moment-expansion} in terms of the $a_{n,m}$, finding
that a moment of order $r$ will be expressed as a linear combination of
$a_{n,m}$'s with $n\leq r$. Moreover, upon tracing over $k$ of the $r$
indices, $m$ will be constrained to be $\leq r-k$. An even/odd number of
indices corresponds to $m$ (and thus $n$) being even/odd.

The lowest order invariants like the mass, the angularities with $a=2,4$ 
and planar flow have the following expression in terms of the Zernike coefficients $a_{n,m}$:
\begin{align*}
\frac{\mJ^2}{E_J^2} &=\frac{\pi}{6}R^2\left(a_{2,0}+3 a_{0,0}\right),\\ 
\tau_{-2} &=\frac{\pi}{240}R^4\left(a_{4,0}+5 a_{2,0}+10 a_{0,0}\right),\\
\tau_{-4} &=\frac{\pi}{4480}R^6\left(a_{6,0}+7 a_{4,0}+21 a_{2,0}+35 a_{0,0}\right),\\
\left(1-\Pf\right)\frac{\mJ^4}{E_J^4} &=\frac{\pi^2}{36}R^4\left(a_{2,2}^2+ a_{2,-2}^2\right).
\end{align*}

In optics, the lowest order coefficients have been given names. In
particular $a_{1,\pm1}$ is the tilt, $a_{2,0}$ is the defocus, $a_{2,\pm
2}$ are the $0^{\circ}$ and $45^{\circ}$-astigmatism (respectively),
$a_{3,\pm1}$ is the coma, and $a_{4,0}$ is the spherical aberration. The
optical analogy may provide us with additional geometrical understanding
of what is being probed by the jet shapes defined in the previous
sections. 

\section{Analysis of $\epsilon I_2 I_4$}
\label{epsI2I4}

We now try to gain some intuition regarding one of the new observables found in section~\ref{first-few-orders},
\mbox{$\cO \equiv 2 \epsilon_{ij} I_{ik} I_{jkmm}$}. Note that any observable
which includes an odd number of $\epsilon_{ij}$ symbols is a
pseudo-scalar with respect to parity on the detector plane. It therefore
vanishes for any energy distribution that is symmetric under reflection
through an arbitrary axis.

Let $I_4'$ be the $I_4$ moment traced once, namely $I_{kkij}$. Both $I_2$
and $I_4'$ are real, symmetric matrices, so we can write them in terms of
the Pauli matrices,
\begin{align*}
  I_2 &= I_{2,0} \frac{\sigma^0}{2} 
  + I_{2,1} \frac{\sigma^3}{2}
  + I_{2,2} \frac{\sigma^1}{2} \ec \\
  I_4' &= I_{4,0} \frac{\sigma^0}{2} 
  + I_{4,1} \frac{\sigma^3}{2}
  + I_{4,2} \frac{\sigma^1}{2} \ed
\end{align*}
The observable can then be written as
\begin{align*}
  \cO &= 2 \epsilon_{ij} I_{ik} I_{jkmm}
  = 2 \trace ( I_2 \epsilon I_4' )
  = \epsilon_{ij} I_{2,i} I_{4,j} \ed
\end{align*}
Note that $\cO$ doesn't depend on the $\sigma^0$ components (the trace). Treating
$I_{2,i},I_{4,j}$ as $d=2$ vectors, we see that
\begin{align}
  \cO = \vec{I}_2 \times \vec{I}_4 \ec
  \label{O}
\end{align}
where the product is the cross product in $d=2$ which produces a
pseudo-scalar.

To get further insight, let us compute the components of these vectors. We do this for a general
moment $I_{2k}'$ with all indices traced except two. It is expanded in
Pauli matrices just like $I_2,I_4'$ above, and we denote its
corresponding $d=2$ vector by $I_{2k}^i$. We write the result using polar
coordinates,
$x_1 = r \cos(\phi)$, $x_2 = r \sin(\phi)$.
\begin{align*}
  I_{2k}^1 &= \trace (I_{2k}' \sigma^3) = I_{2k,11}' - I_{2k,22}'
  = \int_0^R dr \int_0^{2\pi} d\phi \, 
  \en(r,\phi) \, r^{2k+1} \cos(2\phi) \ec\\
  I_{2k}^2 &= \trace (I_{2k}' \sigma^1) = 2 I_{2k,12}'
  = \int_0^R dr \int_0^{2\pi} d\phi \, 
  \en(r,\phi) \, r^{2k+1} \sin(2\phi) \ed
\end{align*}

Plug the Zernike expansion \eqref{zexpansion} of the energy distribution
in $\vec{I}_{2k}$.  The $\phi$ integral picks out the $m=\pm2$ modes, and
the radial integral picks up a linear combination of $a_{2k,\pm2},
\ldots, a_{2,\pm2}$, with $+$ sign for the cosine and $-$ sign for the
sine. Specializing for the cases $k=1,2$ we find:
\begin{align*}
  \vec{I}_2 &= \frac{\pi}{6} R^{2} \begin{pmatrix}
    a_{2,2} \\
    a_{2,-2}
  \end{pmatrix} \ecq
  \vec{I}_4 = \frac{\pi}{40} R^{4}\begin{pmatrix}
    a_{4,2} + 5 a_{2,2} \\
    a_{4,-2}+ 5 a_{2,-2}
  \end{pmatrix} \ed
\end{align*}
The observable \eqref{O} can now be written as
\begin{align}
  \cO &= \frac{\pi^2}{240} R^6\det \begin{pmatrix}
    a_{2,2}  &  a_{4,2} \\
    a_{2,-2}  & a_{4,-2}
  \end{pmatrix} \ec
  \label{Ofinal}
\end{align}
where we have dropped the piece in $\vec I_4$ which is proportional to
$\vec I_2$ because of the usual properties of the determinant.

This shows that, in this expansion of $\en(x)$, $\cO$
depends only on the $m=\pm2$ Fourier modes for the angular part and on two specific combinations
of the even, $n=2,4$ Zernike modes. 

Finally, let us define the normalized observable
\begin{align*}
  \cO_\text{n} &= \frac{\vec{I}_2 \times \vec{I}_4}
  {|\vec{I}_2| |\vec{I}_4|} 
  \ec
\end{align*}
which assumes values in the range $[-1,1]$. This observable is maximal
when $\vec{I}_2$ and $\vec{I}_4$ are orthogonal, and vanishes when
$\vec{I}_2$, $\vec{I}_4$ are linearly dependent. This happens, for
instance, when the energy distribution is invariant under reflection
through some axis. Indeed, in that case we may rotate this axis to
coincide with the $x_1$ direction, following which $a_{n,-2}=0$ since all
the antisymmetric Fourier modes vanish. This agrees with the fact that
this observable is a pseudo-scalar.

Moreover, $I_2$ and $I_4'$ differ only by a different weighing of the
inner/outer part of the jets (due to the additional $r^2$ in
$I_4'$). A symmetric matrix defines a characteristic ellipsoid (in the
case of $I_2$ it is the ellipsoid of inertia in the detector plane). A
non-zero $\vec I_2 \times \vec I_4$ determines by how much the principal axes
of the ellipsoid rotate when we give extra weight to the outer portion of the jet.
(In the language of optics it would correspond to measuring the relative
orientation of the lowest and higher order astigmatism.) With this
intuition it is also trivial to see that $\cO_\text{n}$ vanishes when all
the particles lie at the same distance from the jet axis.
 
\begin{figure}[htb]
\begin{center}
\includegraphics[width=.6\textwidth]{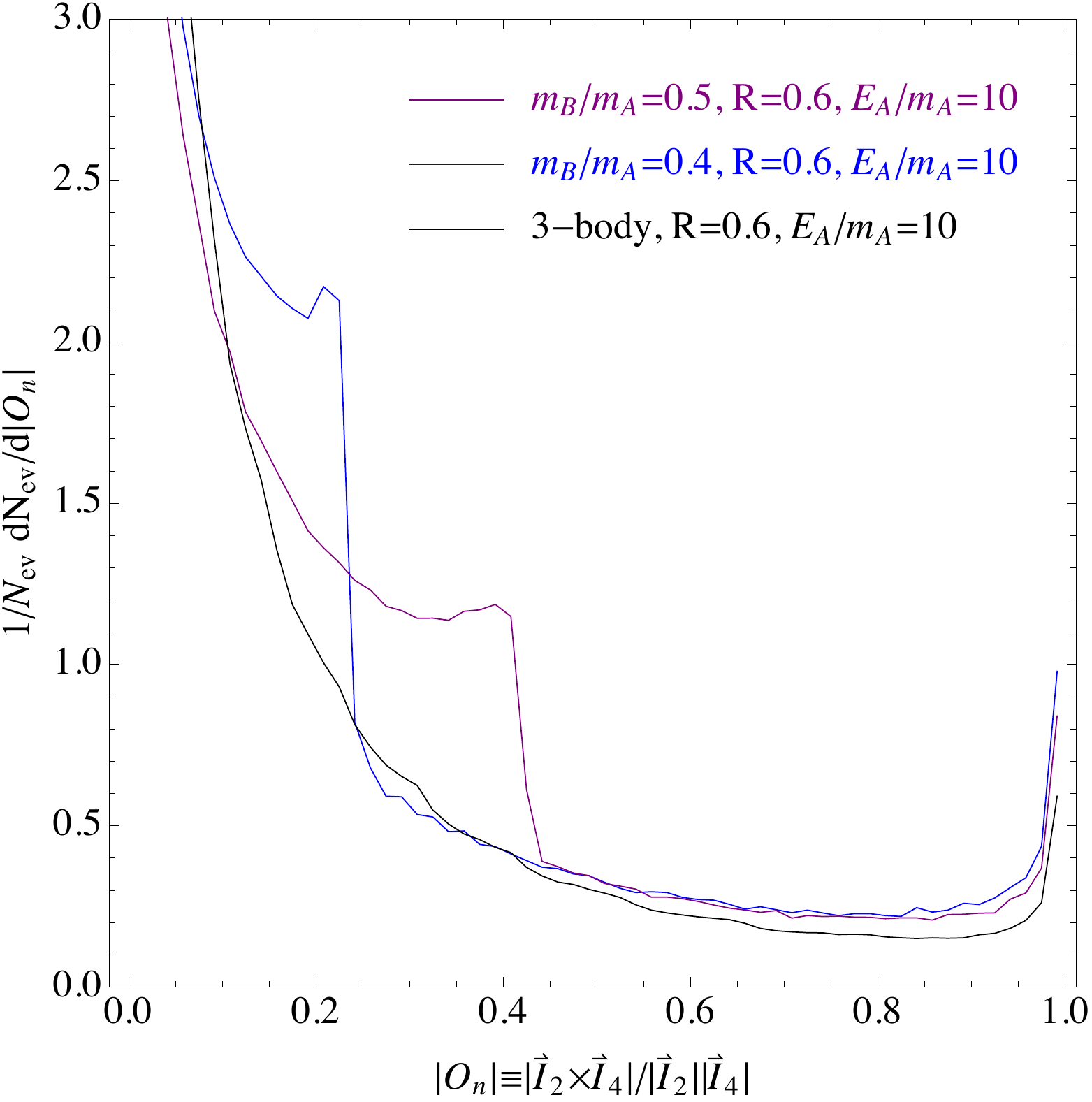}
\caption{The $\cO_\text{n}$ normalized distribution for various values of $m_B/m_A$, the mother particle boost was fixed to $E_A/m_A=10$.}
\label{fig:On}
\end{center}
\end{figure}

We will now briefly study some features of the $\cO_\text{n}$ distribution. While on 
an event-by-event basis $\cO_\text{n}$ will take both positive and negative values, 
in most cases the distribution will be symmetric around zero, yielding $\left<\cO_\text{n}\right>=0$ 
due to its pseudoscalar nature. A necessary condition for $\left<\cO_\text{n}\right>\neq0$ 
is that the substructure of the jet is controlled by a parity violating interaction. 
However, the interesting study of whether and when $\left<\cO_\text{n}\right>$ is non-zero 
is beyond the scope of this work; here we will merely present distributions for $|\cO_\text{n}|$ 
in a few characteristic kinematic scenarios.
Since this observable vanishes for two-particle configurations, in Fig.~\ref{fig:On} we present 
the distributions for three-particle final states. We include both the case of a 
pure one-to-three decay and the case of a three-body decay of particle A via two subsequent 
two-body decays, with an on-shell intermediate particle B (as would be in the case of top decays). 
In the latter case one can see the sharp edge, whose position is just determined by the ratio
\footnote{We checked that it is insensitive to the values of $R$ and of the jet energy $E_J$, 
if varied within reasonable ranges.} $m_B/m_A$.

\section{Discussion}
\label{discussion}

While the study presented in this paper is directly applicable to
$e^+e^-$ machines, let us briefly discuss its applicability to  hadronic
colliders, such as Tevatron and the LHC. Since the partonic center of
mass system is unknown in a hadronic collision, one clearly has to have a
definition for the moments which is invariant under boosts along the beam
axis. At leading order in the narrow cone approximation
this is trivially achieved by using  $\Delta \eta$ and $\Delta \phi$
(differences in pseudorapidity and in azimuthal angle) for the $x_1$ and
$x_2$ coordinates. In fact $\en(\vec x)$ depends only on $\vec x$ and the
ratios $E_i/E_J$. Since $E=p_T \cosh{\eta}$ one finds that $\en(\vec x)$
is also invariant at leading order in the narrow cone approximation.
Corrections start at $O(\tanh{\eta_J} \Delta \eta)$ and contaminate a
moment of order $r$ with moments of order $k>r$ (due to the additional
powers of $\Delta \eta$). However, being of higher order, they will be
suppressed by powers of the cone size $R$, rendering the contamination
small.

Another difference with the $e^+e^-$ case is in the structure of the intra-jet radiation. Indeed in terms of symmetries the presence of beam remnants which can be color-connected with the jet under study provide a specific direction, common to all events, of breaking of the $\SO(2)$ symmetry. While these effects have been widely studied in the literature in various contexts (see {\it e.g}~\cite{Abdesselam:2010pt,factor_soft,Ellis:2010rwa}), it would still be interesting to see how they are represented in the language developed in this paper. While we will not embark here on a systematic study, we will sketch the analysis and outline a few results.

One can relax the requirement of the $\SO(2)$ invariance and still use
the same formalism described in this paper to study these effects.  In
general there will be a set of directions corresponding to the beam axis,
the directions of other jets in the event, etc. They identify
two-dimensional unit vectors $\vec n_k$ in the detector plane that can be
used to contract the indices of the energy flow moment tensors. To lowest
order and considering for simplicity the case of a single $\vec n$ one
has
\begin{align*}
 & n^i n^j I_{i j}\ec \ \ n_i \epsilon^{i j} I_{jk} n^k\ec\\ 
 & I_{iij}n^j\ec \ \ I_{iij} \epsilon^{jk} n_k \ed
\end{align*}
Note also that in our counting of how many independent observables are present for an $N$-particle system, $\SO(2)$ invariance eliminated one observable. Therefore, while the underlying event and soft radiation from color reconnection can contaminate all the jet shape observables presented in this paper, at leading order there is room for only one observable able to probe these effects via a systematic $\SO(2)$ non-invariance. This observable can be chosen to be any combination of the lowest order contractions such as the ones shown above. In particular it is worth noticing that $I_{iij}n^j$ is closely related to $\vec t \cdot \vec n$ where $\vec t $ is the pull variable defined in \cite{Gallicchio:2010sw}, the only differences being a different power of the cone size.

To conclude, we have presented an order-by-order classification of jet shape variables
for narrow jets with massless constituents. At the first few orders we
encountered familiar variables---jet mass, angularity, and planar
flow---and at higher orders we found new observables.  Specifically, we
proposed several new observables that may be used to characterize 3- and
4-particle jets. While the classification is complete, the formalism we
introduced does not provide much insight into the geometric nature of the
observables (beyond a few simple properties), and for that we need
additional tools.  Expanding the energy distribution in terms of Zernike
polynomials seems especially suited for this purpose, at least for the set of observables analyzed here. 
It will be interesting to analyze other observables in this way, 
and in particular the techniques used for studying $\epsilon I_2 I_4$ can be easily applied to other observables
that involve the moments $I_{2k}'$.
A more thorough study of the properties of these new observables, and the identification
of processes for which they provide discriminating power, is an important topic left to future study.

The fact that the Zernike polynomials, which are commonly used in the
field of optics, arise naturally in our jet shape analysis is obviously
related to the fact that in both cases one is describing some
distribution on a disc. One may hope however that the analogy between
optics and jet substructure is deeper. If that is the case, further
insight into the description of jet energy flow may be gained from the 
well-studied theory of aberrations.

\section*{Acknowledgments} 

We thank Walter Goldberger for useful discussions.  The
work of MP was supported in part by the Director, Office of Science,
Office of High Energy and Nuclear Physics, of the US Department of Energy
under Contract DE-AC02-05CH11231.  GP is the Shlomo and Michla Tomarin
career development chair and supported by the Israel Science Foundation
(grant \#1087/09), EU-FP7 Marie Curie, IRG fellowship, Minerva and
G.I.F., the German-Israeli Foundations, and the Peter \& Patricia Gruber
Award.

\end{document}